# Atomic and Electronic Structures of Correlated $SrRuO_3/SrTiO_3$ Superlattices

Seung Gyo Jeong and Woo Seok Choi[*]

*Department of Physics, Sungkyunkwan University, Suwon 16419, Korea*

Ahmed Yousef Mohamed and Deok-Yong Cho[*]

*IPIT & Department of Physics, Jeonbuk National University, Jeonju 54896, Republic of Korea*

Atomic-scale precision epitaxy of perovskite oxide superlattices provides unique opportunities for controlling the correlated electronic structures, activating effective control knobs for intriguing functionalities including electromagnetic, thermoelectric, and electrocatalytic behaviors. In this study, we investigated the close interplay between the atomic and electronic structures of correlated superlattices synthesized by atomic-scale precision epitaxy. In particular, we employ superlattices composed of correlated magnetic $SrRuO_3$ (SRO) and quantum paraelectric $SrTiO_3$ (STO) layers. In those superlattices, $RuO_6$ octahedral distortion is systematically controlled from 167º to 175º depending on the thickness of the STO layers, also affecting the $TiO_6$ octahedral distortion within the STO layer. Customized octahedral distortion within SRO/STO superlattices in turn modifies the electronic structures of both the Ti and Ru compounds, observed by X-ray absorption spectroscopy. Our results identify the close correlation between atomic lattice and electronic structures enabled by the facile controllability of atomic-scale epitaxy, which would be useful for designing future correlated oxide devices.

## I. INTRODUCTION

**Atomic-scale heterostructuring offer a facile control knob for customizing correlated functional oxides.** Perovskite $ABO_3$ oxides exhibit a variety of emergent functional phenomena such as electronic and magnetic phase transitions, superconductivity, ferroelectricity, (anti)ferromagnetism, multiferroicity, various Hall effects, and topologically non-trivial states [1-8]. Recent technological advances in epitaxy and microscopy let us better access the atomic-scale precision oxide heterostructures for future quantum applications [9]. For corner-sharing $BO_6$ networks within the $ABO_3$ perovskites, the bond length and angle of the $B$-O-$B$ are crucial in determining the strong interplay between charge, spin, lattice, orbital, and topological degrees of freedom which gives birth to the intriguing functionality in oxides [10,11]. Numerous oxide heterostructuring approaches have been proposed to modulate the functionalities, for example, cation substitutions of the $B$-site (or $A$-site), oxygen vacancy engineering, dimensionality control, imposing epitaxial strain, and interfacial octahedral symmetry coupling [9,12-16]. Atomically designed superlattices composed of more than two different correlated oxides are useful to modulate the various degrees of freedom via deliberate heterostructuring. Relevant design parameters of superlattice include individual layer compounds and thicknesses, number of repetitions, and stacking order [17]. This provides practical and facile controllability of octahedral distortion and resultant crystalline symmetry of systems, determining correlated functionality with a possibility of emergent phenomena of the superlattices.

**X-ray abortion spectroscopy (XAS) provides atomic-selective electronic structures appropriate for studying perovskite oxide heterostructures [18-22].** XAS is an element-sensitive technique to detect the local electronic structure of an unoccupied orbital state determined by the specific bonding geometry of specific atoms and their orbital hybridization. By controlling the photon energy of incident X-ray, we selectively detect the resonance peaks indicating an electronic transition from the core level to the unoccupied states with the creation of a core hole. The edge structure of the spectrum contains the

chemical information including the oxidation state and coordination environment helpful for figuring out the modulated electronic correlation.

**Stoner ferromagnetic SrRuO3 is a suitable model system to investigate and understand the correlation between the atomic and electronic structures leading to versatile functionalities,** such as dimensional crossover with phase instability, anomalous and/or topological Hall effects, electrocatalyst, THz phonon emission, spin-phonon coupling, and modulation of Weyl fermions [1-3,13,15,23-27]. A variety of SRO heterostructures have been suggested to engineer the $RuO_6$ octahedral distortion via stoichiometry and/or epitaxial strain control of a single SRO film. However, most conventional approaches usually accompany unintended side effects originating from dissimilar substrates, strain relaxation, cation vacancy-induced electronic change, dislocation, and electronic reconstruction.

**In this paper, we demonstrate an atomic-scale heterostructure approach for achieving octahedral distortion control and a systematic modulation of local electronic structure in a correlated oxide superlattice.** To overcome the extrinsic effects in the conventional SRO heterostructure, we use the correlated oxide superlattice composed of ferromagnetic metallic SRO and nonmagnetic insulating $SrTiO_3$ (STO). Here, the identical *A*-site ion (Sr) and the highly suppressed charge transfer between the SRO and STO layers lead to atomically sharp chemical and electronic interfaces [4]. The periodic superlattice structures amplify the experimental signals from atomically thin SRO layers, enabling the conventional characterization by using X-ray diffraction (XRD) and XAS. By atomically controlling the STO layer thickness within SRO/STO superlattices, the octahedral distortion can be controlled systematically, despite the same thickness and stoichiometry of the SRO layer. The customized octahedral distortions for both the $TiO_6$ and $RuO_6$ octahedra modify the collective electronic structures, which is experimentally observed by XAS measurement in this study.

## II. EXPERIMENTS AND DISCUSSION

To control the octahedral distortion in SRO/STO superlattices, we precisely synthesized the epitaxial oxide superlattice composed of 8 atomic unit cells (u.c. ~0.4 nm) of SRO and $y$ u.c. of STO layers with 10 repetitions ([8|$y$] superlattice) on (001)-oriented single crystal STO substrates by using pulsed laser epitaxy [3,4,22,25-27]. We pretreated the STO substrate using buffered HF and annealed it at 1000°C under atmospheric conditions for the atomically flat surfaces of the substrate. We confirmed the typical step-terrace structure on STO substrate by using atomic force microscopy. We utilized a KrF excimer laser (248 nm; IPEX-868, LightMachinery) with 1.5 J/cm$^2$ of fluence and 5 Hz of repetition rate to ablate stoichiometric ceramic SRO and STO targets. We used 750°C and 100 mtorr of oxygen partial pressures for obtaining the stoichiometric SRO and STO layers. We verified the atomically controlled periodicity of superlattice by utilizing a high-resolution XRD with Cu K-$\alpha_1$ (PANalytical).

We performed the XAS measurements at the 16A1 beamline of the Taiwan Light Source in the fluorescence yield (FY) mode for the Ru $L_3$-edge spectra, and the 2A beamline of the Pohang Light Source in the total electron yield (TEY) mode for the O $K$-edge and Ti $L_{2,3}$-edge spectra. XAS spectra were obtained at room temperature with π-polarization, in which the electrical field direction of the X-rays is perpendicular to the sample plane. To obtain the X-ray linear dichroism (XLD) for Ru $L_3$-edge, we rotated the samples with respect to the incoming X-rays by an incidence angle, $\theta_i = 0°$ (beam normal) and $\theta_i = 70°$ (inclined). Whereas the spectra taken at $\theta_i = 0°$ contain horizontal XAS responses ($I_{x,y}$) only, those taken at $\theta_i = 70°$ contain a combination of vertical ($I_z$) and horizontal ($I_{x,y}$) XAS responses as $\cos^2 70° \times I_{x,y} + \sin^2 70° \times I_z$. The $x$, $y$, and $z$ directions in XAS polarization are the $[100]_{pc}$, $[010]_{pc}$, and $[001]_{pc}$ directions of the thin film, respectively, where $[hkl]_{pc}$ denotes the crystallographic direction in perovskite pseudo-cubic notation.

**XRD $\theta$-$2\theta$ measurements confirm the atomic-scale precision control of [8|y] superlattices with different y, as shown in Fig. 1.** XRD $\theta$-$2\theta$ results of [8|y] superlattices show expected superlattice Bragg peaks with well-defined Pendellösung fringes (Fig. 1a), indicating high-quality samples with atomically sharp interfaces and surfaces. The separation between the superlattice peaks increases with decreasing $y$ (dotted lines), corresponding to the decrease in supercell thickness. The thickness of superlattice films was obtained by Bragg's law, $\frac{\lambda}{2}(\sin\theta_n - \sin\theta_{n-1})^{-1}$, where $\lambda$ and $\theta_n$ are the X-ray wavelength for Cu K$\alpha_1$ (0.154 nm) and the angle position of the $n$th-order superlattice peak, respectively. The estimated total thickness of superlattices are 60.50, 54.00, 50.00, 39.24, and 34.66 nm for the $y$ = 8, 6, 4, 2, and 1 superlattices, respectively. These results represent the well-controlled atomic structures of the SRO/STO superlattices. The asterisk (*) denotes the STO substrate peaks and the XRD $\theta$-$2\theta$ data of the SRO single film (~19 nm thick) was included for comparison.

**The atomically controlled STO thickness within the superlattice determines the orthorhombicity ($a_o/b_o$) of the SRO layer within the [8|y] superlattices [22].** We performed the off-axis XRD $\theta$-$2\theta$ scans around (204) STO plane with different $\phi$ (Figs. 1b and S1), to characterize $a_o/b_o$ of SRO single film and superlattices. When $\phi$ = 0º, 90º, 180º, and 270º, the Bragg diffraction peaks shown in Fig. 1b correspond to (44–4)$_o$, (260)$_o$, (444)$_o$, and (620)$_o$ plane [12,28]. Note that $(hkl)_o$ denotes the crystallographic plane for the orthorhombic SRO thin film as schematically shown in Fig. 1 [29]. For the SRO single film, whereas the Bragg peaks of $\phi$ = 0º and 180º appear at the same position, those of $\phi$ = 90º and 270º appear at asymmetric positions indicating an orthorhombic structure with the $a_o/b_o$ value deviating from one. For the [8|y] superlattices, the separation of peak position between the $\phi$ = 90º and 270º configurations decreases systematically with increasing $y$, represented by the decrease of $a_o/b_o$, as shown in Fig. S1. Figure 1c summarizes $a_o/b_o$ values of SRO single film ($y$ = 0) and [8|y] superlattices, as a function of $y$.

Assuming that the change in the bond length between Ru and O ions is nearly identical, which is the case for most transition metal oxides, we estimate the orthorhombic tilt angle ($\theta_t$) as $\theta_t = 180° - 2\cos^{-1}(b_o/a_o)$ [29,30] (Right panel of Fig. 1c and d schematically defines the $a_o$, $b_o$, and $\theta_t$ in the orthorhombic RuO$_6$ octahedra.). Fig. 1d shows that the 165° of $\theta_t$ for SRO single film (comparable to 160° of $\theta_t$ for bulk SRO [29,30]) systematically increases up to 175° for the $y = 8$ superlattice. This result proves that $\theta_t$ of the SRO layer can be effectively modulated by the thickness of adjacent cubic STO layers within the superlattices [3,22], while maintaining the intrinsic properties, e.g., stoichiometry and thickness, of the SRO layers.

**XAS is employed to scrutinize the electronic structures of the SRO layer under the customized octahedral distortion.** Figure 2a schematically shows the experimental configuration of XAS for the SRO/STO superlattices with two different detection modes, i.e., FY for Ru $L_3$-edge and TEY for O $K$- and Ti $L_{2,3}$-edge. Figure 2b exemplarily shows the schematic of the XAS excitation process at the Ru $L_3$-edge (Ru $2p_{3/2}$ to $4d$ transition near a photon energy of 2840 eV) and Ti $L$-edge (Ti $2p_{3/2}$ (or $2p_{1/2}$) to $3d$ transition near 460 eV (or 466 eV)). The probing depth of the Ru $L$-edge XAS is expected to be in the order of microns so that the signals originate throughout the superlattice. Meanwhile, the probing depth of O $K$- and Ti $L$-edge XAS (TEY) is only a few nanometers so that the signals originate predominantly from the superlattice rather than the STO substrate.

XAS at O $K$-edge reflects the unoccupied O $2p$ state that is hybridized with transition metal ions' orbitals. Figure 2c shows the XAS spectra at O $K$-edge for SRO single film and [8|$y$] superlattices. The vertical dotted lines indicate the two peak positions at ~528 and ~530 eV of the unoccupied O $2p$ states hybridized with Ru $4d$ and Ti $3d$ states, respectively. The following peaks above 530 eV are related to oxygen hybridization with Sr $4d$ (533.5–538.0 eV), and Ru $5sp$/Ti $4sp$ (538–547 eV) states [21]. For the [8|$y$] superlattices, two peaks are clearly visible, of which their intensities strongly depend on $y$. With

increasing $y$, **the intensity of Ru-O hybridization (528 eV) systematically decreases while that of Ti-O hybridization (530 eV) significantly increases, manifesting the changes in the abundance of Ru and Ti ions.** On the other hand, for the SRO film (20 nm-thick), only the Ru-O hybridization peak is observed due to the limited probing depth.

Figure 3a shows the Ti $L_{2,3}$-edge XAS spectra of [8|$y$] superlattice with different $y$'s in the TEY mode. The first and the second peaks ($L_3$-edge) can be attributed to Ti $2p_{3/2}\rightarrow 3d$-$t_{2g}$ and 3d-$e_g$ transitions, respectively, while the third and the fourth peaks ($L_2$-edge) can be to Ti $2p_{1/2}\rightarrow 3d$-$t_{2g}$ and 3d-$e_g$ transitions, respectively. Since the $Ti^{4+}$ ions have no $3d$ electrons in their ground state, the electron correlation effect inside the $Ti^{4+}$ ion is negligible. Thus, the energy difference between the first and second peaks (or the third and fourth peaks) can be utilized as a measure of the difference in the effective crystal field strength between Ti $t_{2g}$ and $e_g$ manifolds. The effective crystal field strength ($\Delta$), i.e., the energy difference of the two adjacent peaks in the $L_3$-edge spectrum is 2.3 eV for $y = 1$, while it slightly but systematically increases to 2.4 eV for $y = 8$. The value for $y = 8$ is consistent with the reported value for cubic STO single crystal [31]. Although the actual lattice structure of the STO monolayer would be extremely difficult to characterize, the 0.1 eV decrease as $y$ decreases from 8 to 1, corroborates that the crystal field effect becomes weaker, plausibly because of the distortion in the $TiO_6$ local structure.

Figure 3b shows the atomic picture describing the $y$-dependence in $\Delta$. This scenario is further supported by the fact that the charge transfer between SRO and STO layers is negligibly small and the in-plane lattice parameters of the SRO and STO layers within the superlattice are nearly identical to the STO substrate; The former has been confirmed by the XAS Ti $L_{2,3}$-edge exhibiting $Ti^{4+}$ oxidation state for all the [8|$y$] superlattices and the latter has been confirmed by both XRD reciprocal space map and scanning transmission electron microscopy [3,4,25]. This result suggests that the STO layer thickness within the superlattices affects the lattice structures of not only the SRO (octahedral distortion evidently shown in Fig. 2) but also the STO layers themselves.

**Atomically controlled orthorhombic distortion in SRO/STO superlattices also determines the Ru 4$d$ orbital state and its anisotropy, as shown in Figure 4.** Figures 4a and 4b show XAS at Ru $L_3$-edge with $\theta_i = 0°$ and 70° for the SRO film and the [8|$y$] superlattices. The main doublets from Ru $2p_{3/2}$ → 4$d$ $t_{2g}$ (~2839.5 eV) and $e_g$ (~2842.0 eV) states barely show a noticeable evolution in peak energies with increasing $y$. This suggests robustness in the Ru$^{4+}$ valence states for both SRO film and the superlattices as expected [22]. This confirms that the electronic structure of SRO/STO superlattice is dominantly determined by structural modification not extrinsic effects including charge transfer, defects, nor anti-site disorders. To verify the modulated $\theta_t$-dependent orbital occupation and polarization, we analyze the XLD signal, namely, the difference between horizontal and vertical polarization components in the XAS intensity (XLD = $I_{x,y} - I_z$), as shown in Figure 4c. The $I_z$ ($I_{x,y}$) of Ru $t_{2g}$ states is proportional to the number of unoccupied states in $d_{xz,yz}$ states (sum of 1/2$d_{xy}$ and 1/2$d_{xz,yz}$ states). Hence, negative and positive signals in XLD spectra at the Ru $t_{2g}$ region are related to the $d_{xz,yz}$ and $d_{xy}$ orbital states, respectively, consistent with the previous report on the SRO/STO superlattices [22]. We note that XLD spectra at Ru $e_g$ states show no meaningful trend for $y$, and in turn to $\theta_t$. However, as $y$ increases ($\theta_t$ decreases), the negative XLD signal from $d_{xz,yz}$ orbital disappears, and the positive XLD signal from $d_{xy}$ becomes enhanced. This implies a decrease in occupancy in $d_{xy}$ states due to the structural modifications of SRO/STO superlattices. It is noteworthy that the XLD peak position at $d_{xy}$ states slightly increases (and becomes broader) when $\theta_t$ decreases, suggesting additional Ru $t_{2g}$ splitting between $d_{xz,yz}$ and $d_{xy}$ states. Figure 4d schematically shows the possible scenario of the electronic structure evolution as a function of $\theta_t$ within the superlattices. We propose that the Ru $t_{2g}$ splitting increases and partially leads to the enhancement of the unoccupied $d_{xy}$ state due to the decrease of $\theta_t$ within the SRO/STO superlattices.

## III. CONCLUSION

In conclusion, we precisely control the octahedral distortion in orthorhombic SRO heterostructures to engineer the correlated electronic structures of SRO/STO superlattices. By controlling the STO layer thickness within the superlattices, the Ru-O-Ru distortion angle in the adjacent 8 u.c. SRO layer increases systematically from 167º to 175º. Precisely adjusted octahedral distortion in SRO/STO superlattice determines the electronic structure of both transition metal Ti and Ru compounds within the $TiO_6$ and $RuO_6$ octahedra respectively, as is confirmed by atomic-selective XAS. These results suggest a versatile tunability of atomic-scale heterostructuring to customize the strongly correlated electron systems.

## ACKNOWLEDGEMENT

This work was supported by the Basic Science Research Program through the National Research Foundation of Korea (NRF-2021R1A2C2011340 and 2022R1C1C2006723). D.-Y.C is supported by the research fund of Jeonbuk National University in 2021.

## REFERENCES

[1] S. W. Cho *et al.*, Tailoring topological Hall effect in $SrRuO_3$/$SrTiO_3$ superlattices. Acta Mater. **216**, 117153 (2021).
[2] Z. Fang *et al.*, The Anomalous Hall Effect and Magnetic Monopoles in Momentum Space. Science **302**, 92-95 (2003).
[3] S. G. Jeong *et al.*, Symmetry-Driven Spin-Wave Gap Modulation in Nanolayered $SrRuO_3$/$SrTiO_3$ Heterostructures: Implications for Spintronic Applications. ACS Appl. Nano Mater. **4**, 2160-2166 (2021).
[4] S. G. Jeong *et al.*, Phase Instability amid Dimensional Crossover in Artificial Oxide Crystal. Phys. Rev. Lett. **124**, 026401 (2020).
[5] K. T. Kang *et al.*, Ferroelectricity in $SrTiO_3$ epitaxial thin films via Sr-vacancy-induced tetragonality. Appl. Surf. Sci. **499**, 143930 (2020).
[6] M. Kim *et al.*, Superconductivity in $(Ba,K)SbO_3$. Nat. Mater. **21**, 627-633 (2022).
[7] J. Wang *et al.*, Epitaxial $BiFeO_3$ Multiferroic Thin Film Heterostructures. Science **299**, 1719-1722 (2003).

[8] J. Yue *et al.*, Anomalous transport in high-mobility superconducting SrTiO$_3$ thin films. Sci. Adv. **8**, eabl5668 (2022).
[9] P. D. C. King *et al.*, Atomic-scale control of competing electronic phases in ultrathin LaNiO$_3$. Nat. Nanotechnol. **9**, 443-447 (2014).
[10] Y. Tokura & N. Nagaosa. Orbital Physics in Transition-Metal Oxides. Science **288**, 462-468 (2000).
[11] H. Y. Hwang *et al.*, Emergent phenomena at oxide interfaces. Nat. Mater. **11**, 103-113 (2012).
[12] A. Vailionis, W. Siemons & G. Koster. Room temperature epitaxial stabilization of a tetragonal phase in ARuO$_3$ (A=Ca and Sr) thin films. Appl. Phys. Lett. **93**, 051909 (2008).
[13] S. A. Lee *et al.*, Enhanced electrocatalytic activity via phase transitions in strongly correlated SrRuO$_3$ thin films. Energy Environ. Sci. **10**, 924-930 (2017).
[14] D. G. Schlom *et al.*, Elastic strain engineering of ferroic oxides. MRS Bull. **39**, 118-130 (2014).
[15] D. Kan *et al.*, Tuning magnetic anisotropy by interfacially engineering the oxygen coordination environment in a transition metal oxide. Nat. Mater. **15**, 432-437 (2016).
[16] J. M. Rondinelli, S. J. May & J. W. Freeland. Control of octahedral connectivity in perovskite oxide heterostructures: An emerging route to multifunctional materials discovery. MRS Bull. **37**, 261-270 (2012).
[17] R. Ramesh & D. G. Schlom. Creating emergent phenomena in oxide superlattices. Nat. Rev. Mater. **4**, 257-268 (2019).
[18] J. Yano & V. K. Yachandra. X-ray absorption spectroscopy. Photosynth. Res. **102**, 241 (2009).
[19] Z. Liao *et al.*, Large orbital polarization in nickelate-cuprate heterostructures by dimensional control of oxygen coordination. Nat. Commun. **10**, 589 (2019).
[20] X. Liu *et al.*, Interfacial charge-transfer Mott state in iridate–nickelate superlattices. Proc. Natl. Acad. Sci. U.S.A. **116**, 19863-19868 (2019).
[21] H. Jeong *et al.*, Thickness-dependent orbital hybridization in ultrathin SrRuO$_3$ epitaxial films. Appl. Phys. Lett. **115**, 092906 (2019).
[22] S. G. Jeong *et al.*, Propagation Control of Octahedral Tilt in SrRuO$_3$ via Artificial Heterostructuring. Adv. Sci. **7**, 2001643 (2020).
[23] H. I. Seo *et al.*, Crystalline symmetry-dependent magnon formation in the itinerant ferromagnet SrRuO$_3$. Phys. Rev. B **103**, 045104 (2021).
[24] G. Koster *et al.*, Structure, physical properties, and applications of SrRuO$_3$ thin films. Rev. Mod. Phys. **84**, 253-298 (2012).
[25] S. G. Jeong *et al.*, Unconventional interlayer exchange coupling via chiral phonons in synthetic magnetic oxide heterostructures. Sci. Adv. **8**, eabm4005 (2022).
[26] S. G. Jeong *et al.*, Spin–phonon coupling in epitaxial SrRuO$_3$ heterostructures. Nanoscale **12**, 13926-13932 (2020).
[27] S. G. Jeong, A. Seo & W. S. Choi. Atomistic Engineering of Phonons in Functional Oxide Heterostructures. Adv. Sci. **9**, 2103403 (2022).
[28] D. Kan & Y. Shimakawa. Strain Effect on Structural Transition in SrRuO$_3$ Epitaxial Thin Films. Cryst. Growth Des. **11**, 5483-5487 (2011).
[29] S. H. Chang *et al.*, Thickness-dependent structural phase transition of strained SrRuO$_3$ ultrathin films: The role of octahedral tilt. Phys. Rev. B **84**, 104101 (2011).
[30] A. T. Zayak, X. Huang, J. B. Neaton & K. M. Rabe. Structural, electronic, and magnetic properties of SrRuO$_3$ under epitaxial strain. Phys. Rev. B **74**, 094104 (2006).
[31] W. Fan *et al.*, Evolution of element-specific electronic structures in alkaline titanates. AIP Adv. **9**, 065213 (2019).

Figure Captions.

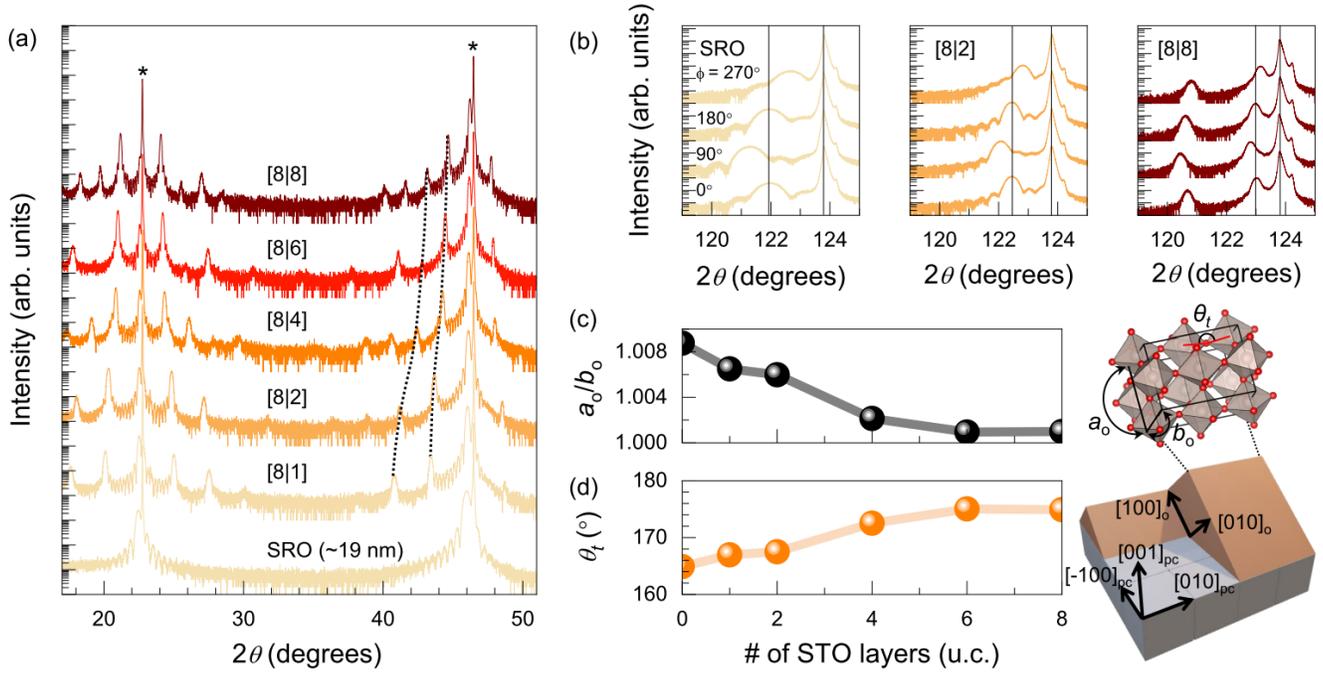

Fig. 1. Structural characterization of atomically designed SRO/STO superlattices: **a** XRD $\theta$-$2\theta$ scans for SRO single film and [8|y] superlattices. **b** off-axis XRD scans around (204) plane of STO with different $\phi$ angles for SRO single film, [8|2], and [8|8] superlattices. **c** $a_o/b_o$ values obtained by off-axis XRD scans and **d** estimated tilt angle $\theta_t$ as a function of $y$. The right panel of **c,d** shows a schematic illustration of $RuO_6$ octahedra with $\theta_t$. The $[hkl]_{pc}$ and $[hkl]_o$ indicates the crystallographic direction in perovskite pseudo-cubic and orthorhombic SRO notation, respectively.

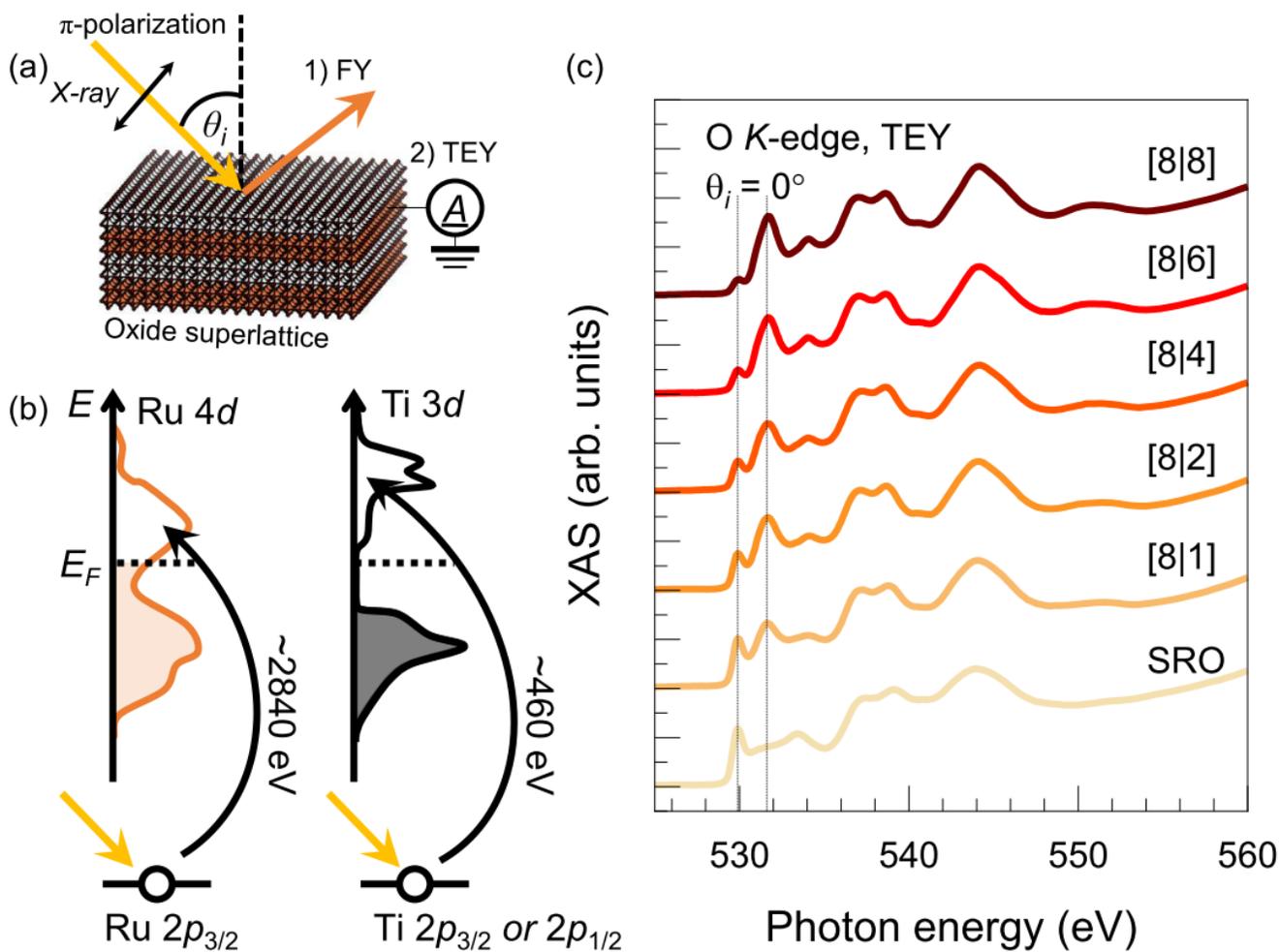

Fig. 2. **a** Schematic representation of XAS configuration for oxide superlattice with two different detection modes (fluorescence yield and total electron yield). **b** atomic-selective excitation process of XAS for Ru and Ti ions by controlling photon energy of X-ray. **c** O *K*-edge XAS spectra of [8|*y*] superlattice taken at $\theta_i = 0°$. The vertical lines indicate the peak positions related to Ru-O and Ti-O hybridization as guides to the eye.

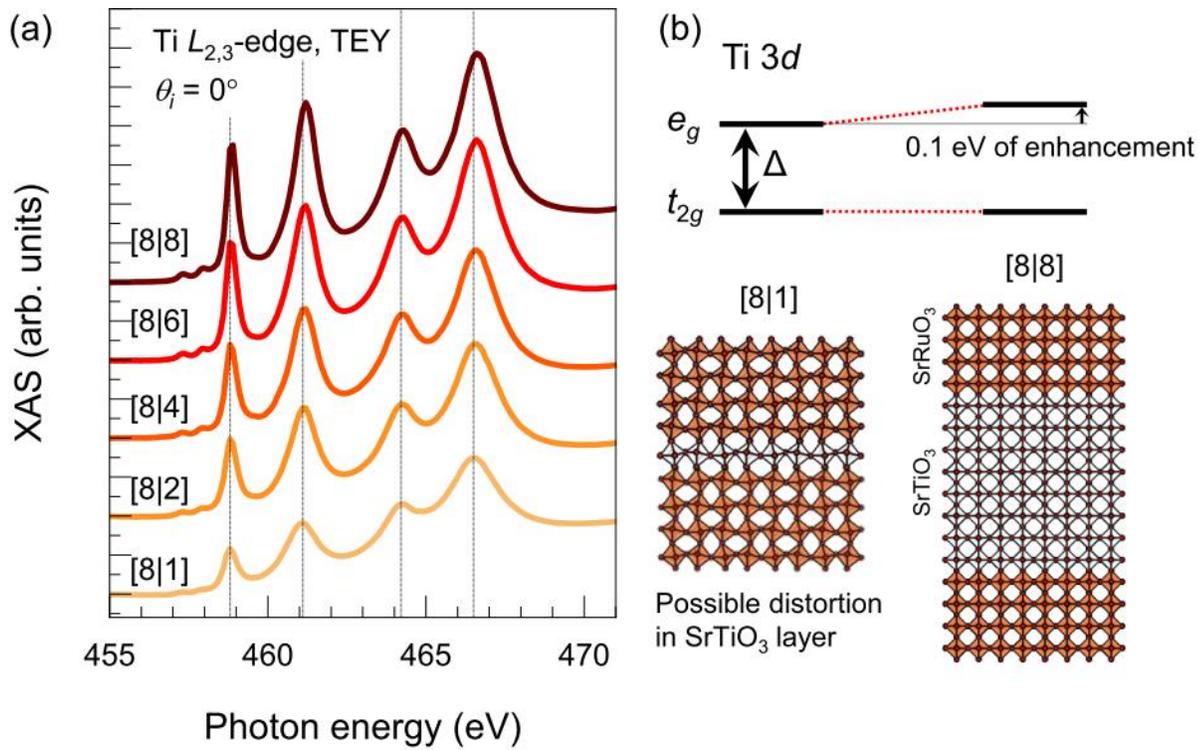

Fig. 3. **a** Ti $L_{2,3}$-edge XAS results of the [8|y] superlattices taken at $\theta_i = 0°$. The vertical lines are guides to the eye. **b** Schematic diagram of the Ti 3d orbital states with a possible y-dependent octahedral distortion in the STO layer.

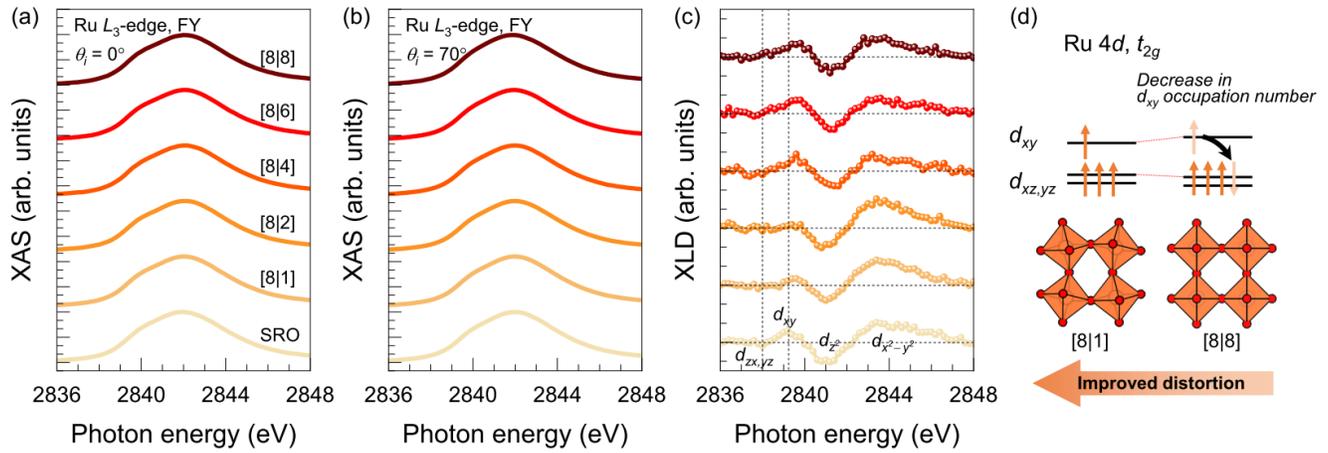

Fig. 4. **a** XAS spectra at Ru $L_3$-edge for the SRO film and the [8|y] superlattices with **a** $\theta_i = 0°$ and **b** $\theta_i = 70°$. **c** XLD = $I_{x,y} - I_z$. The vertical lines are guides to the eye. **d** Schematic illustration of possible Ru orbital states within the SRO/STO superlattice as a function of $\theta_t$.